\documentclass[12pt,preprint]{aastex}

\begin{document}

\title{Spectroscopy of $i^\prime$-Dropout Galaxies with 
       an NB921-Band Depression in the Subaru Deep Field}

\author{
          T. Nagao            \altaffilmark{1, 2},
          N. Kashikawa        \altaffilmark{2},
          M. A. Malkan        \altaffilmark{3},
          T. Murayama         \altaffilmark{4},
          Y. Taniguchi        \altaffilmark{4}, \\
          K. Shimasaku        \altaffilmark{5},
          K. Motohara         \altaffilmark{6},
          M. Ajiki            \altaffilmark{4},
          Y. Shioya           \altaffilmark{4}, 
          K. Ohta             \altaffilmark{7}, \\
          S. Okamura          \altaffilmark{5, 8}, and
          M. Iye              \altaffilmark{2}
}

\altaffiltext{1}{INAF -- Osservatorio Astrofisico di Arcetri,
         Largo E. Fermi 5, 50125 Firenze, Italy; tohru@arcetri.astro.it}
\altaffiltext{2}{National Astronomical Observatory of Japan,
         2-21-1 Osawa, Mitaka, Tokyo 181-8588, Japan}
\altaffiltext{3}{Department of Astronomy, University of California 
         at Los Angeles, P. O. Box 951547, Los Angeles, CA 90095-1547}
\altaffiltext{4}{Astronomical Institute, Graduate School of Science,
         Tohoku University, Aramaki, Aoba, Sendai 980-8578, Japan}
\altaffiltext{5}{Department of Astronomy, Graduate School of Science,
         University of Tokyo, 7-3-1 Hongo, Bunkyo, Tokyo 113-0033, Japan}
\altaffiltext{6}{Institute of Astronomy, Graduate School of Science,
         University of Tokyo, 2-21-1 Osawa, Mitaka, Tokyo 181-0015, Japan}
\altaffiltext{7}{Department of Astronomy,  Graduate School of Science,
         Kyoto University, Kitashirakawa, Sakyo, Kyoto 606-8502, Japan}
\altaffiltext{8}{Research Center for the Early Universe, Graduate School
         of Science, University of Tokyo, Tokyo 113-0033, Japan}

\begin{abstract}
We report new spectroscopy of two star-forming
galaxies with strong Ly$\alpha$ emission at $z=6.03$ and $z=6.04$ 
in the Subaru Deep Field.
These two objects are originally selected as $i^\prime$-dropouts
($i^\prime - z^\prime > 1.5$) showing an interesting
photometric property, the ``NB921 depression''. The
$NB921$-band (centered at 9196${\rm \AA}$) magnitude is significantly
depressed with respect to the $z^\prime$-band magnitude.
The optical spectra of these two objects exhibit asymmetric 
emission-lines at $\lambda_{\rm obs} \sim 8540 {\rm \AA}$ and
$\sim 8560 {\rm \AA}$, suggesting that these objects are Ly$\alpha$
emitters at $z \sim 6$.
The rest-frame equivalent widths of the Ly$\alpha$ emission of 
the two objects are 94${\rm \AA}$ and 236${\rm \AA}$;
the latter one is the Ly$\alpha$ emitter with the largest Ly$\alpha$ 
equivalent width at $z \gtrsim 6$ ever 
spectroscopically confirmed.
The spectroscopically measured Ly$\alpha$ fluxes of these two objects
are consistent with the interpretation that the NB921 depression is
caused by the contribution of the strong Ly$\alpha$ emission
to the $z^\prime$-band flux. 
Most of the NB921-depressed $i^\prime$-dropout objects are thought
to be strong Ly$\alpha$ emitters at $6.0 \lesssim z \lesssim 6.5$;
Galactic L and T dwarfs and NB921-dropout galaxies at $z > 6.6$ do
not dominate the NB921-depressed $i^\prime$-dropout sample.
Thus the NB921-depression method is very useful for finding 
high-$z$ Ly$\alpha$ emitters with a large Ly$\alpha$ equivalent
width over a large redshift range, $6.0 \lesssim z \lesssim 6.5$.
Although the broadband-selected sample at $z \sim 3$ contains
only a small fraction of objects with a Ly$\alpha$ equivalent width 
larger than $100 {\rm \AA}$, the $i^\prime$-dropout sample of the
Subaru Deep Field contains a much larger fraction of such strong 
Ly$\alpha$ emitters.
This may imply a strong evolution of the Ly$\alpha$ equivalent 
width from $z \gtrsim 6$ to $z \sim 3$.
\end{abstract}


\section{INTRODUCTION}

Observational studies of galaxy evolution generally 
rely on the investigation of statistical properties such
as luminosity functions, correlation functions,
distribution functions of mass and emission-line 
equivalent widths, as functions of redshift from 
$z \sim 0$ to the highest redshift, i.e., $z>6$.
This is one of the reasons why we should explore the 
highest-$z$ universe and construct a highest-$z$ galaxy sample. 
Recently, two alternative methods are often used 
to find high-$z$ galaxies. One is to search for objects 
with a Lyman-break feature due to intergalactic absorption
seen as a strong photometric discontinuity in the rest-frame
UV stellar continuum spectrum (e.g., Steidel et al. 1996a, 
1996b; Madau et al. 1996; Lowenthal et al. 1997).
The other is to search for objects with a strong 
Ly$\alpha$ emission caused by massive stars using narrow
passband filters (see for a review, Taniguchi et al. 2003). 
The high-$z$ galaxies found in such ways are called 
``Lyman-break galaxies (LBGs)'' and ``Ly$\alpha$ emitters 
(LAEs)'', respectively.
These high-$z$ galaxy surveys are now bringing us a sample of
galaxies at $z>6$ or the end of ``the cosmic dark age'' (e.g., 
Miralda-Escud\'{e} 2003; Djorgovski 2004).
Recent narrow-band surveys pick up LAEs at $z\sim6.6$
through an atmospheric window at $\sim9200{\rm \AA}$ (e.g.,
Hu et al. 2002; Kodaira et al. 2003; Rhoads et al. 2004;
Taniguchi et al. 2005), 
and recent broad-band surveys pick up ``$i^\prime$-dropout
galaxies'' as candidates of galaxies at $z\sim6$, which are 
characterized by a very red $i^\prime - z^\prime$ color due 
to a Lyman-break spectral feature (e.g., 
Stanway et al. 2003; Bouwens et al. 2003, 2004;
Dickinson et al. 2004; Bunker et al. 2004).
In order to investigate the nature of such high-$z$
galaxy populations, follow-up spectroscopy is
crucially important.

Recently, Nagao et al. (2004) reported the discovery of an
interesting object among 48 $i^\prime$-dropout 
galaxies detected in the Subaru Deep Field 
(SDF\footnote{The SDF project consists of
multi-color deep imaging observations by Suprime-Cam
(Miyazaki et al. 2002) and spectroscopic observations
by FOCAS (Kashikawa et al. 2002), boarded on the 
Subaru Telescope (Kaifu et al. 2000; Iye et al. 2004).
SDF is centered on (13$^h$24$^m$38$\fs$9, 
+27$\arcdeg$29$\arcmin$25$\farcs$9) (J2000.0).
See Kashikawa et al. (2004) and Taniguchi et al. (2005)
for more details.}; Kashikawa et al. 2004).
The spectrum of this object, SDF J132440.6+273607, 
which has been obtained during the SDF
follow-up spectroscopy campaign (see Taniguchi et al.
2005 for details), shows a very
strong Ly$\alpha$ emission at 8909${\rm \AA}$, i.e.,
$z=6.33$, whose rest-frame equivalent width is as large as 
$EW_0 = 130{\rm \AA}$. However, this object does not show
strong UV stellar continuum that is expected for
$i^\prime$-dropout galaxies.
Thus the flux detected in the $z^\prime$ image is largely 
($\sim$40\%) attributed to the Ly$\alpha$ emission, giving
a very red $i^\prime - z^\prime$ color.
This discovery is important because
only $\sim$ 1\% of broad-band selected LBGs at $z\sim3$ 
show strong Ly$\alpha$ emission with $EW_0$(Ly$\alpha$) $>$
100${\rm \AA}$ (Shapley et al. 2003). The finding of 
SDF J132440.6+273607 among such a small sample of
$i^\prime$-dropout galaxies may suggest that an 
$i^\prime$-dropout sample contains strong LAEs much more 
frequently than a LBG sample at a redshift of $\sim$3.

To study this issue further, we focus on the 48 
$i^\prime$-dropout objects found in SDF. This is because
we have not only deep $i^\prime$ and $z^\prime$ images
but also a narrow-band image of NB921 [($\lambda_{\rm c}$,
$\Delta \lambda_{\rm FWHM}$) = (9196${\rm \AA}$,
132${\rm \AA}$); the transmission curves of these filters are
shown in Figure 1]. This enables us to obtain information
about the SED of the $i^\prime$-dropout galaxies. 
As for SDF J132440.6+273607, 
due to the strong Ly$\alpha$ emission in the $z^\prime$ band
(but out of the NB921 band), a significant ``NB921 depression'' 
is seen; i.e., $z^\prime - NB921 = -0.54$ . 
Interestingly, among the 48 $i^\prime$-dropout objects in SDF,
there are seven other objects with a NB921 
depression with significance larger than 2$\sigma$
of the sky noise (note that the significance level of
NB921 depression for SDF J132440.6+273607 is slightly below 
2$\sigma$). If all of these 7 objects are really
similar objects (i.e., high-$z$ galaxies with
very large $EW_0$(Ly$\alpha$)), SDF J132440.6+273607 
would not be exceptional in the $i^\prime$-dropout 
population.  This might suggest a strong redshift evolution
of the $EW_0$(Ly$\alpha$) distribution from $z\sim6$
to $z\sim3$ although it should be kept in mind that
$z\sim3$ LBGs and $z\sim6$ $i^\prime$-dropout galaxies 
are not selected by the same criteria.

In this paper, we report the results of our new spectroscopy
of two $i^\prime$-dropout objects with a significant NB921 
depression. Throughout this paper, we adopt a cosmology with 
($\Omega_{\rm tot}$, $\Omega_{\rm M}$, $\Omega_{\Lambda}$)
=(1.0, 0.3, 0.7) and $H_0$ = 70 km s$^{-1}$ Mpc$^{-1}$.
We use the AB photometric system for optical magnitudes.

\section{OBSERVATIONS}

The SDF spectroscopic follow-up campaign has been
carried out with DEIMOS (Faber et al. 2003) on the Keck II 
telescope, on 2 nights in April 2004. During this observing run, 
we obtained spectra of $i^\prime$-dropout objects which 
were selected by the following four criteria; 
(1) $i^\prime - z^\prime > 1.5$,
(2) $z^\prime < 26.1$ (i.e., above 5$\sigma$),
(3) $B > 28.5$ (below 3$\sigma$), and
(4) $R_{\rm C} > 27.8$ (below 3$\sigma$).
The adopted $i^\prime - z^\prime$ color criterion is the same
as that adopted for  $i^\prime$-dropouts by
Stanway et al. (2003).
In this observing run, we observed two NB921-depressed
objects, SDF J132426.5+271600 and
SDF J132442.5+272423 among seven NB921-depressed
$i^\prime$-dropout objects in the SDF. Here the objects that
satisfy the fifth criterion are classified as
NB921-depressed $i^\prime$-dropout objects;
(5) $z^\prime - NB921$ showing a depression whose significance
    level is larger than 2$\sigma$ of the sky noise.
In Figure 2, the photometric properties of the 48
$i^\prime$-dropout objects in SDF are shown on a
$z^\prime - NB921$ versus $z^\prime$ color-magnitude diagram
and a $z^\prime - NB921$ versus $i^\prime - z^\prime$
color-color diagram. The photometric properties of the
targeted two objects are given in Table 1. See Kashikawa 
et al. (2004) for details of the SDF imaging observations.

The spectra of SDF J132426.5+271600 and SDF J132442.5+272423
were obtained on 23 April 2004 (UT). The 830 lines/mm grating 
and the GG495 order-sorting filter were used.
The resulting wavelength coverage was $\sim 6000 {\rm \AA} - 9600 {\rm \AA}$,
with a dispersion of 0.47 ${\rm \AA}$/pixel.
The adopted slit width was 1.0 arcsec, which gave a spectral
resolution of $R \sim 3600$ at 8500${\rm \AA}$, which is 
measured by the widths of atmospheric OH emission lines.
The total integration time was 7929 sec for SDF J132426.5+271600
and 7058 sec for SDF J132442.5+272423.
The weather was clear and photometric, and
the typical seeing size was 0.7 -- 1.0 arcsec during the observation.
We also obtained spectra of a spectrophotometric
standard star Feige 110 (Oke 1990) for the flux calibration.
The obtained data were reduced by the DEEP2 DEIMOS data
reduction software, the spec2d pipeline\footnote{
The data reduction pipeline was developed at UC Berkeley with 
support from NSF grant AST-0071048.}.

\section{RESULTS}

In Figure 3, the sky-subtracted optical spectra are plotted
with a typical sky spectrum.
Both objects show a single strong emission line,
with peak wavelengths of 8540${\rm \AA}$ and 8560${\rm \AA}$
for SDF J132426.5+271600 and SDF J132442.5+272423, respectively.
No continuum emission is detected for both objects.
Both emission lines show a clear asymmetric profile; i.e.,
a sharp decline on the blue side and a prominent tail on
the red side of the line. Spiky features redward of
the emission line of SDF J132426.5+271600 seem to be
the residuals of the sky emission. 
The redward tails extend to $\sim 8564 {\rm \AA}$ 
and $\sim 8577 {\rm \AA}$ for SDF J132426.5+271600 and 
SDF J132442.5+272423 respectively.
To quantify the asymmetry of the lines,
we measured the flux ratio between $f_{\rm red}$ and 
$f_{\rm blue}$, where $f_{\rm red}$ is the flux at wavelengths
longer than the peak while $f_{\rm blue}$ is 
that at wavelengths shorter.
The measured ratios are $2.5 \pm 0.4$ and $1.6 \pm 0.2$ 
for SDF J132426.5+271600 and SDF J132442.5+272423 respectively,
which are significantly larger than unity,
although the spiky noise redward of the emission line of SDF 
J132426.5+271600 would affect the ratio. % of $f_{\rm red}$/$f_{\rm blue}$.
This is consistent with the interpretation that the detected
emission line is Ly$\alpha$, not any other
emission line (see Taniguchi et al. 2005). The observed line-shape
asymmetry and the photometric properties strongly suggest 
that SDF J132426.5+271600 and SDF J132442.5+272423 are at 
$z = 6.03$ and $z = 6.04$, respectively.
This interpretation is supported by the fact that no other
emission lines are detected for both objects, because
H$\alpha$, H$\beta$, or [O{\sc iii}]$\lambda$5007 emitters at
$z \sim 0.30$, $z \sim 0.76$, or $z \sim 0.71$ (which are 
possible candidates for line emitters at $\sim 8550 {\rm \AA}$)
should show [O{\sc iii}] at $\sim 6510 {\rm \AA}$,
[O{\sc iii}] at $\sim 8810 {\rm \AA}$ or H$\beta$ at 
$\sim 8310 {\rm \AA}$, respectively. The possibility that
the observed objects are [O{\sc ii}] emitters at $z \sim 1.29$ 
is rejected by the obtained emission-line profiles, because
the expected wavelength separation of the [O{\sc ii}] doublet 
for objects at $z = 1.29$ is $\Delta \lambda \sim 6.4 {\rm \AA}$, 
which should be resolved by our wavelength resolution of $R \sim 3600$. 
Note that the wavelength separation of the emission-line peak and the
spiky feature redward of the emission line of SDF J132426.5+271600 is 
$\sim 8 {\rm \AA}$, rather larger than the separation expected
for the [O{\sc ii}] doublet from an object at $z = 1.29$.
We also note that the $i^\prime - z^\prime$ color of 
foreground galaxies at $z<1.3$ is expected to be bluer than
1.5 mag, even when the SED of passively evolved elliptical-type
galaxies is assumed (e.g., Stanway et al. 2003).
We thus conclude that the observed two objects are strong LAEs at
$z = 6.03$ and $z = 6.04$.

Adopting these redshifts, the N{\sc v}$\lambda$1240 emission
would be expected at $\sim 8680 {\rm \AA}$ 
if these two objects are AGNs at $z \sim 6$.
However, no such emission-line features are
seen in this wavelength for either object (Figure 3).
The 3$\sigma$ upper limit for the N{\sc v}$\lambda$1240 emission
is $9.5 \times 10^{-18}$ ergs s$^{-1}$ cm$^{-2}$ and
$9.3 \times 10^{-18}$ ergs s$^{-1}$ cm$^{-2}$ for
SDF J132426.5+271600 and SDF J132442.5+272423 respectively.
However, we cannot reject the possibility that these
two objects are AGNs only by these weak constraints on the 
N{\sc v}$\lambda$1240 fluxes. Indeed it is known that
some high-$z$ narrow-line radio galaxies show very weak
N{\sc v}$\lambda$1240 emission compared with Ly$\alpha$
(e.g., De Breuck et al. 2000; Nagao et al. 2005a). In order to examine this issue
further, follow-up spectroscopy in the near-infrared is required to 
see some high-ionization emission lines such as 
C{\sc iv}$\lambda$1549 attributed to AGN. Despite this 
uncertainty, however, we regard these two observed objects as
star forming galaxies but not AGNs in the following analysis
and discussion.

The Ly$\alpha$ fluxes of SDF J132426.5+271600 and 
SDF J132442.5+272423 are $(3.6 \pm 0.3) \times 10^{-17}$ 
ergs s$^{-1}$ cm$^{-2}$ and $(4.5 \pm 0.3) \times 10^{-17}$ 
ergs s$^{-1}$ cm$^{-2}$, respectively, giving Ly$\alpha$ 
luminosities of $(1.5 \pm 0.1) \times 10^{43}$ ergs s$^{-1}$ 
and $(1.8 \pm 0.1) \times 10^{43}$ ergs s$^{-1}$
(where the slit losses are not taken into account).
The measured emission-line widths of SDF J132426.5+271600 and 
SDF J132442.5+272423 are $12.0 {\rm \AA}$ and $6.8 {\rm \AA}$ 
in full-width at half maximum (FWHM), corresponding to
410 km s$^{-1}$ and 220 km s$^{-1}$ where the instrumental 
broadening effect is quadratically corrected. The estimated 
line widths at zero intensity (FWZI) are 
$29.5 {\rm \AA}$ and $22.0 {\rm \AA}$, corresponding to 
1030 km s$^{-1}$ and 770 km s$^{-1}$, respectively.
The measured spectroscopic properties of the two observed
objects are summarized in Table 2.

While the $NB921$-band flux is attributed only to
the stellar UV continuum emission, the $z^\prime$-band fluxes 
of the two observed objects apparently contain much Ly$\alpha$ 
flux. The estimated fractions of the $z^\prime$-band flux contributed
by Ly$\alpha$ are $\sim 25 \%$ and $\sim 46 \%$ 
for J132426.5+271600 and SDF J132442.5+272423, respectively.
The UV continuum flux densities derived from the $NB921$
magnitudes are $5.5 \times 10^{-20}$ ergs s$^{-1}$ cm$^{-2}$
${\rm \AA}^{-1}$ and $2.7 \times 10^{-20}$ ergs s$^{-1}$ 
cm$^{-2}$ ${\rm \AA}^{-1}$, respectively. 
We then estimate the Ly$\alpha$-subtracted $z^\prime$-band 
magnitude, by subtracting the spectroscopically measured 
Ly$\alpha$ flux from the $z^\prime$-band flux. Note that
this procedure may result in an over-subtraction due to
a short-wavelength cutoff of the $z^\prime$-filter transmission.
The estimated magnitudes are $z^\prime_{\rm cor} = 25.7$ and 
$z^\prime_{\rm cor} = 26.6$ for SDF J132426.5+271600 and 
SDF J132442.5+272423, respectively. These magnitudes are nearly 
the same as the $NB921$ magnitudes within the range of $<0.2$ mag.
This result suggests that the NB921 depression is caused
mainly by the contribution of the Ly$\alpha$ flux into
the $z^\prime$-band flux.

For narrow-band selected high-$z$ LAEs,
it is sometimes difficult to estimate $EW_0$(Ly{$\alpha$})
because the measurements of continuum flux on the
low-S/N spectra are generally very hard.
Indeed only upper-limits to the UV continuum flux density
and resulting lower limits to $EW_0$(Ly{$\alpha$}) have been obtained for
such LAEs in most of the previous high-$z$ (i.e., $z \gtrsim 6$) 
LAE studies (e.g., Kodaira et al. 2003; Kurk et al. 2004;
Taniguchi et al. 2005). However, for SDF J132426.5+271600 and 
SDF J132442.5+272423, we can estimate $EW$(Ly{$\alpha$}) rather
straightforwardly by referring to the $NB921$-band flux to 
determine the continuum levels, if we assume a flat UV SED. 
The estimated values of
$EW_{\rm obs}$(Ly{$\alpha$}) are 660 ${\rm \AA}$ and
1659 ${\rm \AA}$, giving $EW_0$(Ly{$\alpha$}) = 94 ${\rm \AA}$
and $EW_0$(Ly{$\alpha$}) = 236 ${\rm \AA}$ for SDF J132426.5+271600 
and SDF J132442.5+272423, respectively. Note that if the 
detected emission lines were [O{\sc ii}]$\lambda$3727, the
rest-frame equivalent widths would be 288${\rm \AA}$ and
724${\rm \AA}$, respectively. These values are implausibly large
for low-$z$ [O{\sc ii}] emitters (e.g., Ajiki et al. 2005),
which also supports our interpretation that the detected 
emission lines are Ly$\alpha$. Here we mention that the derived
$EW_0$(Ly{$\alpha$}) for SDF J132442.5+272423 is the largest one
among the spectroscopically identified galaxies ever known
at $z \gtrsim 6$.

\section{DISCUSSION}

\subsection{The Nature of $NB921$-Depressed $i^\prime$-Dropout Objects}

Although SDF J132426.5+271600 and SDF J132442.5+272423 turned
out to be strong LAEs at $z \sim 6$, it does not necessarily mean
that all of the NB921-depressed $i^\prime$-dropout objects
are similar strong LAEs.
We point out that there are three possibilities
for the NB921-depressed $i^\prime$-dropout objects;
Galactic very late-type stars,
NB921-dropout galaxies at $z > 6.6$, and
strong Ly$\alpha$ emitters at $6.0 \lesssim z \lesssim 6.5$.
To understand the nature of the observed objects selected
by the criterion ``NB921-depressed $i^\prime$-dropout'',
we discuss each possibility below.

\subsubsection{Galactic Late-Type Stars}

Objects selected by $i^\prime$-dropout criterion are not
necessarily high-$z$ galaxies; sometimes Galactic late-type
stars contaminate in $i^\prime$-dropout samples
(e.g., Stanway et al. 2004).
The NB921 depression might be accompanied with such
late-type stars owing to their strong and complex molecular 
absorption spectral features.
Therefore it is worthwhile to examine whether such Galactic
late-type stars can be selected as NB921-depressed 
$i^\prime$-dropout objects before considering the possibility
that they are really very high-$z$ galaxies.
In Figure 4, the photometric properties of the NB921-depressed 
$i^\prime$-dropout objects are compared with those of
Galactic stars on a $z^\prime - NB921$ versus 
$i^\prime - z^\prime$ color-color diagram. Here we use 
Gunn \& Stryker (1983) for photometric properties of Galactic
M type and earlier, and Kirkpatrick (2003) for 
those of L and T dwarf stars. 
This figure shows that Galactic L and T dwarfs exhibit 
red $i^\prime - z^\prime$ color ($i^\prime - z^\prime > 1.5$)
and thus are selected as $i^\prime$-dropout objects, but
none of them exhibit significant NB921 depression.
The bluest $z^\prime - NB921$ color of the L and T
dwarfs is $z^\prime - NB921 \sim +0.1$, which is too red
to reproduce the observed NB921 depressions seen in our
$i^\prime$-dropout subsample.
Although the possibility that Galactic stars are included in
the NB921-depressed $i^\prime$-dropout sample cannot be
completely discarded when taking the photometric uncertainties
into account, Figure 5 suggests that late-type Galactic stars 
are not a dominant population in
the NB921-depressed $i^\prime$-dropout sample.

\subsubsection{NB921-Dropout Galaxies at $z > 6.6$}

We now consider the second possibility that the NB921-depressed 
$i^\prime$-dropout galaxies are ``NB921-dropout galaxies'',
i.e., galaxies at $z > 6.6$ and thus the Lyman-break feature
drops at a longer wavelength than the NB921 band.
See Shioya et al. (2005) for general discussion on 
NB921-dropout galaxies. To see the photometric properties of
the NB921-dropout galaxies quantitatively, theoretically 
calculated $z^\prime - NB921$ color
is plotted as a function of redshift in Figure 5.
The adopted stellar spectrum is created by the stellar population
model GALAXEV (Bruzual \& Charlot 2003) assuming a stellar
metallicity of $Z = 0.02$, a Salpeter initial mass function
($0.1 \leq M/M_\odot \leq 100$), and an exponentially decaying
star-formation history with the timescale of $\tau = 1$ Gyr
and the age of 1 Gyr. % where $SFR(t) \propto$ exp($-t/\tau$).
Although no dust reddening is taken into
account for models shown in Figure 6, a moderate amount of the
reddening does not affect the $z^\prime - NB921$ color
significantly; a dust reddening of $E(B-V)=0.3$ mag changes
the $z^\prime - NB921$ color less than $\pm 0.2$ mag.
Models with $EW_0$(Ly{$\alpha$}) = 
0, 65, 130, and 260 ${\rm \AA}$ are plotted in the figure.
As shown in Figure 5, significant NB921 depressions can be 
caused at the redshift range of $6.7 \lesssim z \lesssim7.2$
due to the combination of the absorption of the $NB921$ flux
by the intergalactic matter and the contribution of
the Ly$\alpha$ contribution into the $z^\prime$-band flux.
However, the $z^\prime$-band flux of galaxies at $z > 6.6$
decreases so significantly that the detection of such
NB921-dropout galaxies is very difficult.
In Figure 6, the expected $z^\prime$ magnitude of galaxies
with an absolute magnitude of $M_{1500} = -20.0$ is shown
as a function of redshift.  The adopted SEDs are the same as 
those in Figure 6. It is clearly shown that the 
$z^\prime$-band flux is decreasing at $z > 6.6$, especially 
for the galaxies with no or weak Ly$\alpha$ emission.
Shioya et al. (2005) predicted the expected number of 
NB921-dropout galaxies in the FOV of the SDF. They
reported that the expected number of NB921-dropout galaxies 
with $z^\prime < 26.1$ mag is $\approx$0.03 if a starburst SED 
without Ly$\alpha$ is assumed and $\approx$1.2 if 
$EW_0$(Ly{$\alpha$}) = 65 ${\rm \AA}$ is assumed 
(here a UV luminosity function at $z \sim 6$ reported by 
Bouwens et al. 2004 is adopted). 
This suggests that the NB921-dropout galaxies
(i.e., galaxies at $z > 6.6$) are at most a minor population
in the NB921-depressed $i^\prime$-dropout sample.

\subsubsection{Strong LAEs at $6.0 \lesssim z \lesssim 6.5$}

For the two LAEs presented in this paper and another LAE
reported by Nagao et al. (2004), the NB921 depression 
of the NB921-depressed $i^\prime$-dropout objects can be 
attributed to the large Ly$\alpha$ contribution in the 
$z^\prime$-band flux if the strong LAE is in
proper redshift range.
As shown in Figure 5, the $z^\prime - NB921$ color is not 
affected by Ly$\alpha$ at $z \lesssim 5.8$ because Ly$\alpha$
is out of the $z^\prime$ band. 
At $6.0 \lesssim z \lesssim 6.5$, the NB921 depression is
caused by a contribution of strong Ly$\alpha$ emission into
the $z^\prime$-band. Indeed we carried out spectroscopy of 
the three such objects in the SDF
(two in this paper and one in Nagao et al. 2004; see Figure 2)
and all of the three objects now turn out to be strong LAEs 
with $EW_0$(Ly{$\alpha$}) $\gtrsim 100 {\rm \AA}$
at the redshift range of $6.0 \lesssim z \lesssim 6.5$.
Since Galactic late-type dwarfs and NB921-dropout galaxies
at $z > 6.6$ are thought not to be a dominant populations 
in the NB921-depressed $i^\prime$-dropout objects,
most of the NB921-depressed $i^\prime$-dropout galaxies seem to
be strong LAEs, and their NB921 depression seems to be 
caused by the large Ly$\alpha$ contribution into the 
$z^\prime$-band flux.

\subsection{NB921-Depression Method to Search for Strong LAEs}

We have shown that the NB921-depression is a powerful search method
for strong LAEs.
%seems interesting in terms of the following viewpoints. 
First, by this method one can search for high-$z$ LAEs with a
large $EW_0$(Ly{$\alpha$}) in a larger redshift range,
$6.0 \lesssim z \lesssim 6.5$, than the traditional NB921-excess
method ($6.5 \lesssim z \lesssim 6.6$).
Although Ajiki et al. (2004) proposed a method of LAE
surveys using an intermediate-band filter in order to
sweep larger volumes than those using the usual 
narrow-band filters (see also Fujita et al. 2003), 
the NB921-depression method can sweep a
much larger volume for strong LAEs.
% than using an intermediate-band filter.
And second, this method picks up LAEs with a very large 
$EW_0$(Ly{$\alpha$}) selectively. 
The LAEs with a large $EW_0$(Ly{$\alpha$}) are highly
interesting because a very young stellar population or
a top-heavy initial-mass function would be expected for
LAEs with $EW_0$(Ly{$\alpha$}) $> 100 {\rm \AA}$
(e.g., Leitherer et al. 1999; Malhotra \& Rhoads 2002).
And more interestingly, a population III stellar cluster
is also expected to show very large
$EW_0$(Ly{$\alpha$}) because it radiates much
more hydrogen-ionizing photons per unit mass than
population I/II stellar clusters (e.g., Schaerer 2002, 2003;
Venkatesan, Tumlinson, \& Shull 2003; Scannapieco,
Schneider, \& Ferrara 2003).
To pursue these interesting high-$z$ populations,
the NB921-depression method is a highly efficient strategy
(see also Nagao et al. 2005b)
and thus the application of this method to other deep
fields will be an excellent way to
investigate high-$z$ populations further.

Our interest in LAEs with a large $EW_0$(Ly{$\alpha$})
naturally leads to the following question: does 
the frequency distribution of $EW_0$(Ly{$\alpha$})
evolve on a cosmological timescale? At $z \sim 3$,
$\approx$1\% of LBGs (i.e., broadband-selected objects) 
show $EW_0$(Ly{$\alpha$}) $> 100 {\rm \AA}$
(Sharpley et al. 2003). On the contrary,
at $z \gtrsim 6$, at least 3 objects among 48 
$i^\prime$-dropout objects in SDF show $EW_0$(Ly{$\alpha$}) 
$> 100 {\rm \AA}$. Note that we are now focusing on
broadband-selected objects; narrow-band selected objects
tend to show much larger $EW_0$(Ly{$\alpha$}) 
(e.g., Malhotra \& Rhoads 2002; Dawson et al. 2004).
Since other 5 NB921-depressed 
$i^\prime$-dropout galaxies would also possibly possess
$EW_0$(Ly{$\alpha$}) $> 100 {\rm \AA}$ as discussed above,
our results suggest that the frequency distribution of 
$EW_0$(Ly{$\alpha$}) evolves significantly from 
$z \gtrsim 6$ to $z \sim 3$,
which may imply the evolution of some properties of
stellar populations such as the mean stellar age or the
initial mass function (see, e.g., Kudritzki et al. 2000;
Malhotra \& Rhoads 2002).
Note that some difference in the $EW_0$(Ly{$\alpha$}) 
frequency distribution is caused by a selection effect,
because LBGs at $z \sim 3$ are selected by two colors while
$i^\prime$-dropout galaxies at $z \gtrsim 6$ are selected by
a $i^\prime - z^\prime$ color alone.
Our criterion for $i^\prime$ dropouts, $i^\prime - z^\prime > 1.5$,
is so strict that we might pick up objects with
a large $EW_0$(Ly{$\alpha$}) selectively.
This issue should be examined more quantitatively by future
complete spectroscopic surveys of $i^\prime$-dropout galaxies.
We should also be aware that we see somewhat different 
ranges of the luminosity functions for the $z \sim 3$ sample and
the $z \gtrsim 6$ sample discussed here. 
While the spectroscopic survey for the $z \sim 3$ LBG
sample by Sharpley et al. (2003) reaches down to 
$R_{\rm AB} \sim 25.5$ mag which corresponds to
$L_{\rm lim} \sim L_\ast (z=3) + 1.0$ (see Steidel et al. 1999),
the SDF $i^\prime$-dropout sample consists of galaxies with
$z^\prime > 25.9$ mag which corresponds to
$L_{\rm lim} \sim L_\ast (z=6) + 0.5$ if adopting the UV
luminosity function at $z \sim 6$ reported by Bouwens et al.
(2004). While the above discussion on the evolution of
the $EW_0$(Ly{$\alpha$}) frequency distribution would not be
affected if emission-line equivalent widths are independent of
luminosity of galaxies as suggested by, e.g., van Dokkum et al. 
(2004), other observations imply that the emission-line 
equivalent widths may depend on the luminosity of galaxies
(e.g., Ando et al. 2004). 
We do not discuss this issue further because this topic is 
beyond the scope of this paper.

\subsection{Star-Formation Rates of $NB921$-Depressed 
            $i^\prime$-Dropout Galaxies}

We now consider the star-formation rates (SFRs) of the
NB921-depressed $i^\prime$-dropout galaxies. The SFRs of
the two NB921-depressed $i^\prime$-dropout galaxies
are obtained by adopting the following relation; 
$SFR$(Ly$\alpha$) = $9.1 \times 10^{-43}$ $L$(Ly$\alpha$)
$M_\odot$ yr$^{-1}$ where $L$(Ly$\alpha$) is in units of
ergs s$^{-1}$ (Kennicutt1998; Brocklehurst 1971). Then
$SFR$(Ly$\alpha$) = 13.4 $M_\odot$ yr$^{-1}$ 
and $SFR$(Ly$\alpha$) = 16.6 $M_\odot$ yr$^{-1}$ for 
SDF J132426.5+271600 and SDF J132442.5+272423, respectively. 
These SFRs are lower limits because no corrections are made 
for possible absorption effects on the Ly$\alpha$ emission.
We can also estimate the SFRs from the luminosity of the UV 
stellar continuum by adopting the following relation;
$SFR({\rm UV}) = 1.4 \times 10^{-28} L_{\nu} M_{\odot} 
{\rm yr}^{-1}$ (Kennicutt 1998). By using the UV continuum 
flux densities measured from the $NB921$ magnitudes,
we obtain $SFR({\rm UV}) =$ 9.9 $M_\odot$ yr$^{-1}$ 
and $SFR({\rm UV}) =$ 4.8 $M_\odot$ yr$^{-1}$ for 
SDF J132426.5+271600 and SDF J132442.5+272423, respectively. 
Here the Salpeter initial mass function (0.1--100 $M_\odot$)
and the flat UV SED are assumed to derive $SFR$(Ly$\alpha$)
and $SFR({\rm UV})$ (see Kennicutt 1998).

The ratios of SFRs estimated by the two methods, 
$SFR$(Ly$\alpha$)/$SFR({\rm UV})$, are significantly higher
than unity in both of the two observed objects, which is
similar to the other NB921-depressed 
$i^\prime$-dropout galaxy reported by Nagao et al. (2004).
This result is interesting because high-$z$ LAEs with smaller 
$EW_0$(Ly{$\alpha$}) tend to show the opposite trend
(i.e., $SFR$(Ly$\alpha$)/$SFR({\rm UV})$ $<$ 1; e.g.,
Hu et al. 2002; Ajiki et al. 2003; Taniguchi et al. 2005),
possibly due to resonance scattering effects on the
Ly$\alpha$ emission. 
As mentioned by Nagao et al. (2004), this systematic difference
in the ratio of $SFR$(Ly$\alpha$)/$SFR({\rm UV})$ between LAEs
with small $EW_0$(Ly{$\alpha$}) and large $EW_0$(Ly{$\alpha$}) 
may be due to a different stellar population because 
large $EW_0({\rm Ly}\alpha)$ ($>100 {\rm \AA}$) is difficult to  
reproduce by continuous star formation with a normal 
initial mass function, which is assumed when we derive the SFRs.
This trend is consistent with the description of Schaerer (2000)
in which $SFR({\rm UV})$ is underestimated
for star-forming galaxies with an age of $< 10^8$ yr
(see also, e.g., Leitherer et al. 1999; Malhotra \& Rhoads 2002).

\section{SUMMARY}

Based on DEIMOS follow-up spectroscopy of objects imaged in the SDF, we
identified two NB921-depressed $i^\prime$-dropout objects
to be strong LAEs at $z \sim 6$.
Taking both this finding and theoretical considerations into
account, we conclude that most of the NB921-depressed
$i^\prime$-dropout objects in the SDF are strong LAEs at
$6.0 \lesssim z \lesssim 6.5$.
The presence of a large fraction of LAEs with a large
$EW_0$(Ly$\alpha$) may imply the evolution of $EW_0$(Ly$\alpha$)
at high redshift.
This has implications for the cosmological evolution of stellar populations,
and the existence of a population III.

\vspace{0.5cm}

The SDF project is one of the investigations promoted by the Subaru
Telescope, which is operated by the National Astronomical 
Observatory of Japan. We thank the staffs of the Subaru Telescope
and of the W. M. Keck Observatory. We also thank M. Doi for his 
contribution to the data acquisition. TN and MA are JSPS fellows.
This research is partly supported by the Japan Society for the
Promotion of Science (JSPS) through Grant-in-Aid 15340059 and
16740118, and by the Ministry of Education, Culture, Sports,
Science and Technology (10044052 and 10304013).
TN and MA are JSPS fellows.

%-------------------------------------------------------------------------

\clearpage
%-----------------------------------------------------
%    Table 1
%-----------------------------------------------------

\begin{deluxetable}{lcc}
\tablenum{1}
% \footnotesize
\tablecaption{Photometric Properties of the Target Objects}
\tablewidth{0pt}
\tablehead{
 \colhead{Object} &
 \colhead{Band} &
 \colhead{Magnitude\tablenotemark{a}} \\
 \colhead{} & 
 \colhead{} & 
 \colhead{(mag)}
}
\startdata
SDF J132426.5+271600 & $B$        & $>28.5$\tablenotemark{b} \\
                     & $R_{\rm C}$& $>27.8$\tablenotemark{b} \\
                     & $i^\prime$ & 27.43 \\
                     & $z^\prime$ & 25.36 \\
                     & $NB921$    & 25.92 \\
SDF J132442.5+272423 & $B$        & $>28.5$\tablenotemark{b} \\
                     & $R_{\rm C}$& $>27.8$\tablenotemark{b} \\
                     & $i^\prime$ & 27.69 \\
                     & $z^\prime$ & 25.74 \\
                     & $NB921$    & 26.71 \\
\enddata
\tablenotetext{a}{AB magnitude measured within a $2\farcs0$ diameter aperture.}
\tablenotetext{b}{3$\sigma$ upper-limit magnitudes.}
\end{deluxetable}

%-----------------------------------------------------
%    Table 2
%-----------------------------------------------------

\begin{deluxetable}{lccccccc}
\rotate
\tablenum{2}
% \tabletypesize{\small}
% \scriptsize
% \footnotesize
\tablecaption{Spectroscopic Properties of the Target Objects}
\tablewidth{0pt}
\tablehead{
 \colhead{Object} &
 \colhead{Redshift} &
 \colhead{$F$(Ly$\alpha$)} &
 \colhead{$L$(Ly$\alpha$)} &
 \colhead{FWHM\tablenotemark{a}} &
 \colhead{FWZI\tablenotemark{a}} &
 \colhead{$f_{\rm red}/f_{\rm blue}$} &
 \colhead{$EW_0$(Ly$\alpha$)}         \\
 \colhead{} &
 \colhead{} &
 \colhead{(10$^{-17}\!$ ergs/s/cm$^{2}$)} &
 \colhead{(10$^{43}\!$ ergs/s)} &
 \colhead{(km/s)} &
 \colhead{(km/s)} &
 \colhead{} &
 \colhead{($\rm \AA$)}
}
\startdata
SDF J132426.5+271600 & 6.03 & 3.6 $\pm$ 0.3 & 
  1.5 $\pm$ 0.1 & 410 & 1030 & 2.5 $\! \pm \!$ 0.4 & 94\\
SDF J132442.5+272423 & 6.04 & 4.5 $\pm$ 0.3 & 
  1.8 $\pm$ 0.1 & 220 &  770 & 1.6 $\! \pm \!$ 0.2 & 236\\
\enddata
\tablenotetext{a}{Corrected for the instrumental broadening.}
\end{deluxetable}
\clearpage

%-----------------------------------------------------
%    Figures
%-----------------------------------------------------

\begin{figure}
\epsscale{0.70}
\plotone{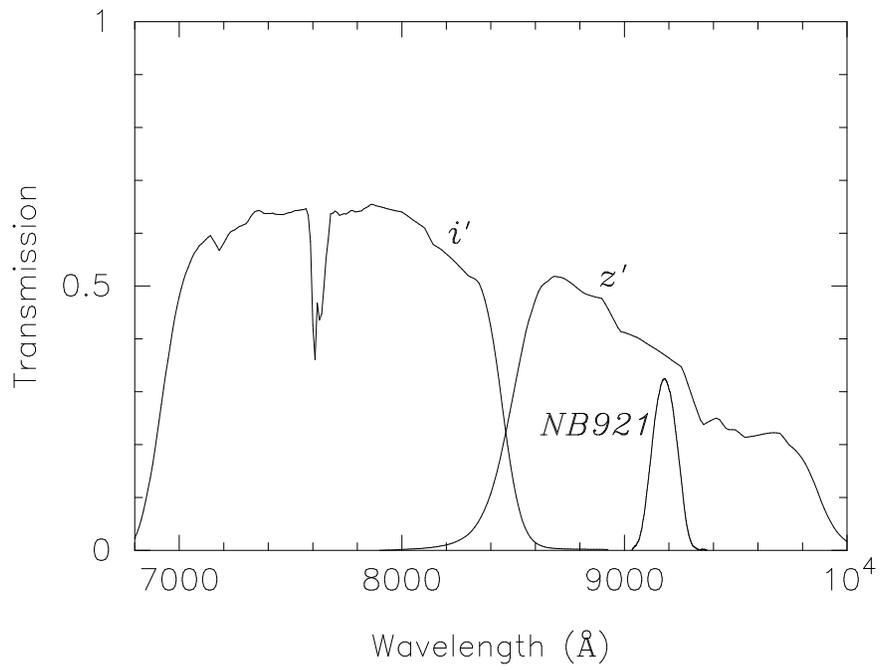}
\caption{
Transmission curves of the $i^\prime$, $z^\prime$ and
$NB921$ filters installed in Suprime-Cam.
The response curve of MIT CCD detectors on Suprime-Cam,
the efficiency of the telescope optics including the mirror
and the prime-focus corrector, and the atmospheric
opacity (sec$z$ = 1.2) are taken into account.
\label{fig1}}
\end{figure}

\begin{figure}
\epsscale{0.90}
\plotone{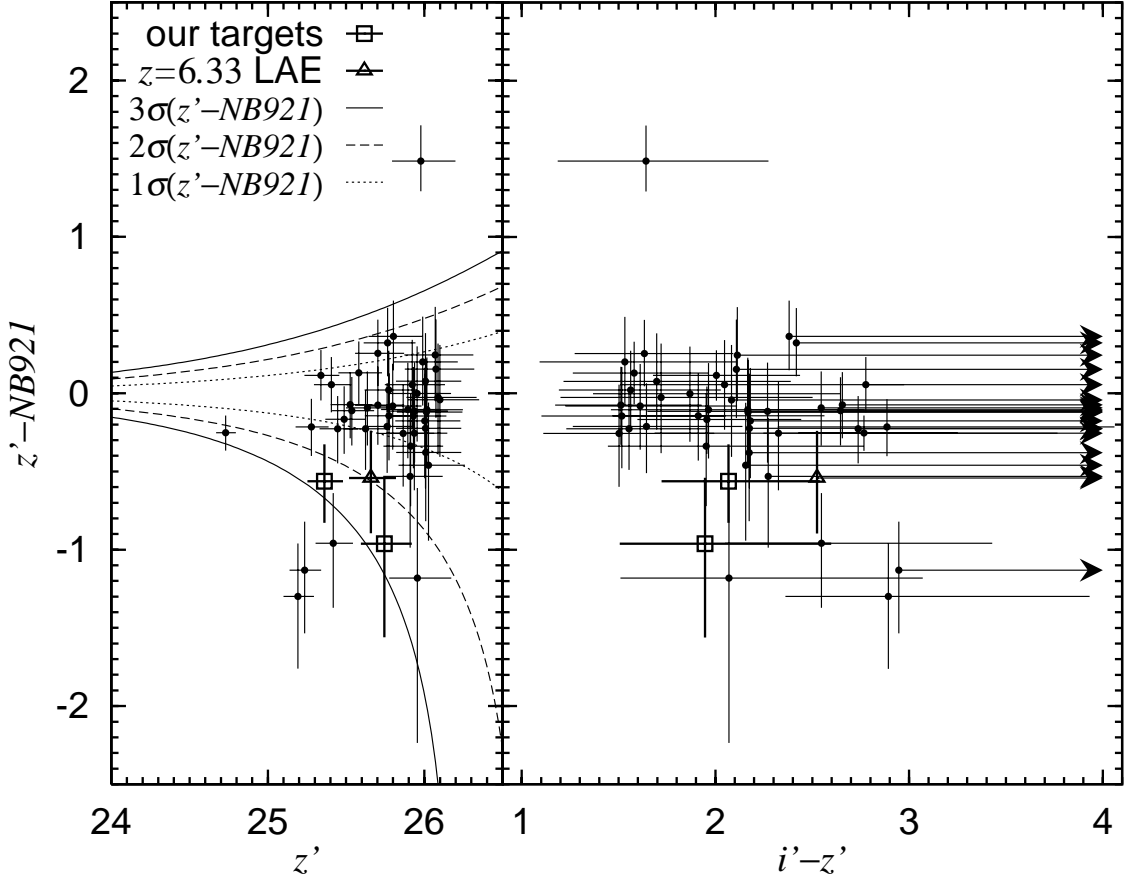}
\caption{
$z^\prime - NB921$ versus $z^\prime$ color-magnitude diagram ($left$)
and $z^\prime - NB921$ versus $i^\prime - z^\prime$ color-color 
diagram ($right$). The 48 $i^\prime$-dropout objects
detected in the SDF are plotted. The two target objects among them
(SDF J132426.5+271600 and SDF J132442.5+272423) are shown by
open squares, and the objects discussed by Nagao et al. (2004)
is shown by a open triangle. In the left panel, 
1$\sigma$, 2$\sigma$, and 3$\sigma$ uncertainties
of the color of $z^{\prime} - NB921$ are shown by
the dotted, dashed, and solid lines, respectively.
\label{fig2}}
\end{figure}

\clearpage%\newpage

\begin{figure}
\epsscale{1.00}
\plotone{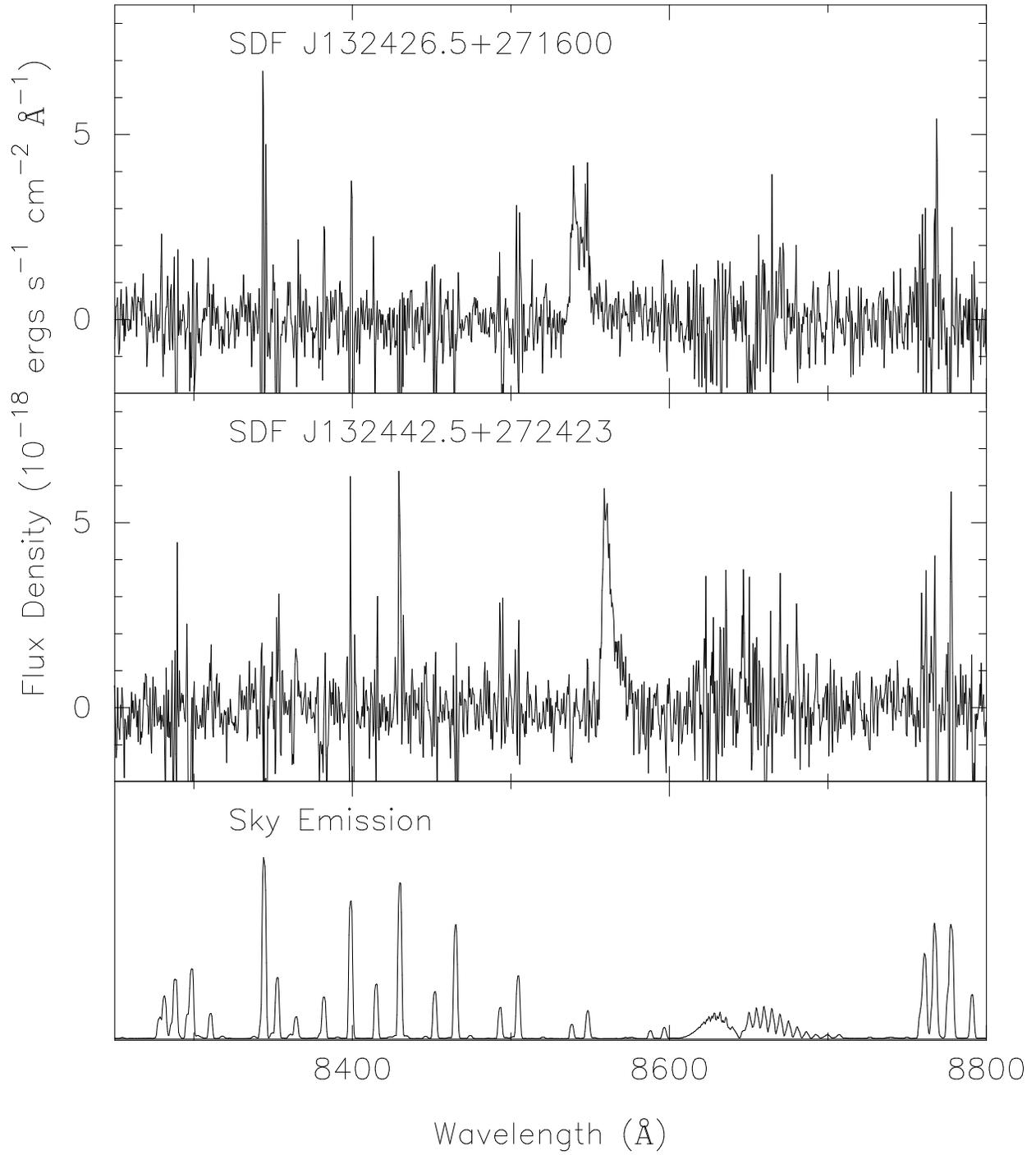}
\caption{
Optical spectra of SDF J132426.5+271600 ($upper$), 
SDF J132442.5+272423 ($middle$) and typical sky emission
($lower$).
\label{fig3}}
\end{figure}

\clearpage%\newpage

\begin{figure}
\epsscale{0.95}
\plotone{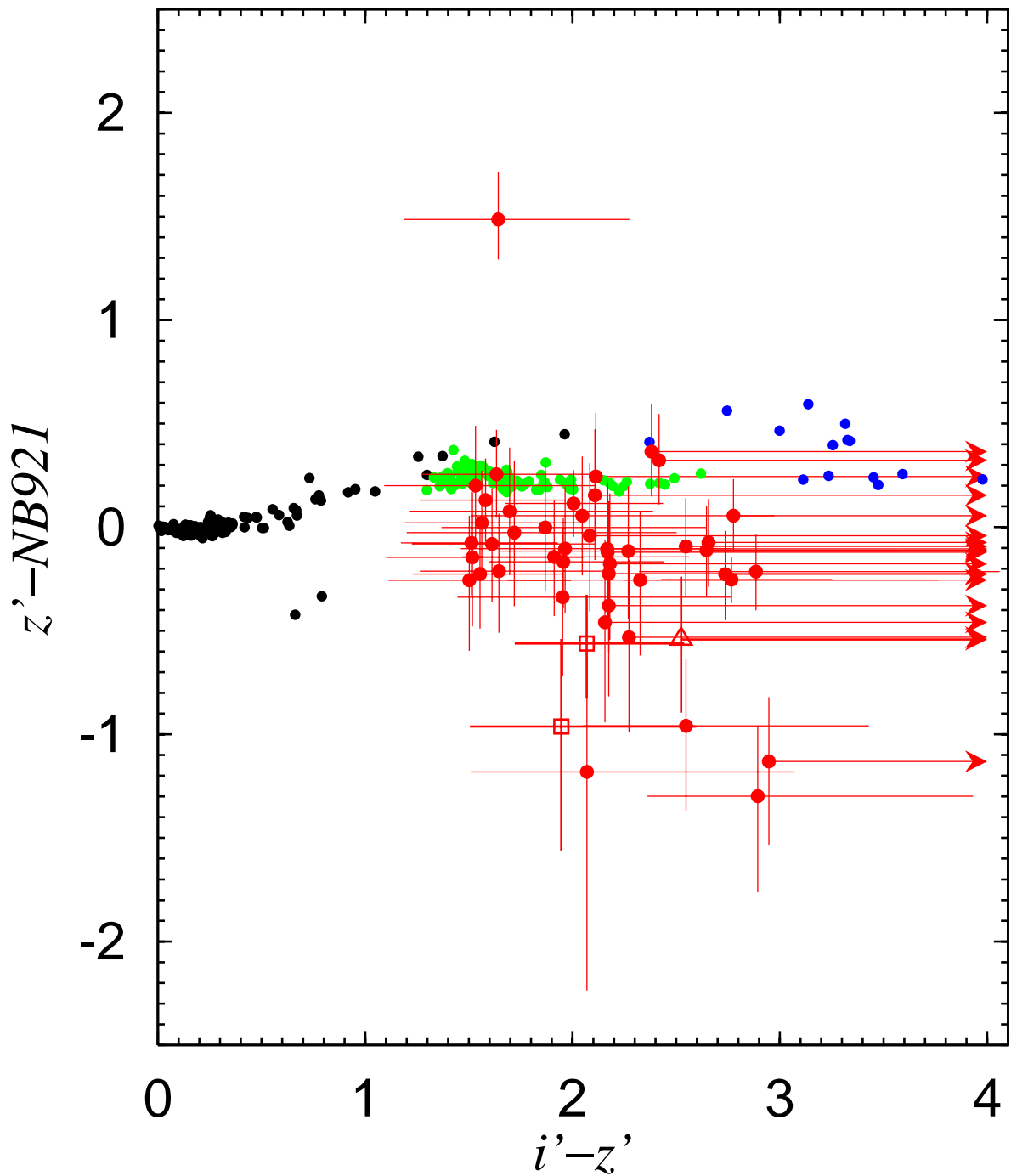}
\caption{
Photometric properties of Galactic stars and observed
NB921-depressed $i^\prime$-dropout objects plotted on
$z^\prime - NB921$ versus $i^\prime - z^\prime$ color-color 
diagram. Black, green, and blue filled circles denote
Galactic M and earlier type stars (Gunn \& Stryker 1983),
L dwarfs and T dwarfs (Kirkpatrick 2003), respectively.
The data of the NB921-depressed $i^\prime$-dropout objects
are shown by red points; among them, the two target objects 
(SDF J132426.5+271600 and SDF J132442.5+272423) are shown by
open squares, and the object discussed by Nagao et al. 
(2004) is shown by a open triangle. 
\label{fig4}}
\end{figure}

\clearpage%\newpage

\begin{figure}
\epsscale{1.00}
\plotone{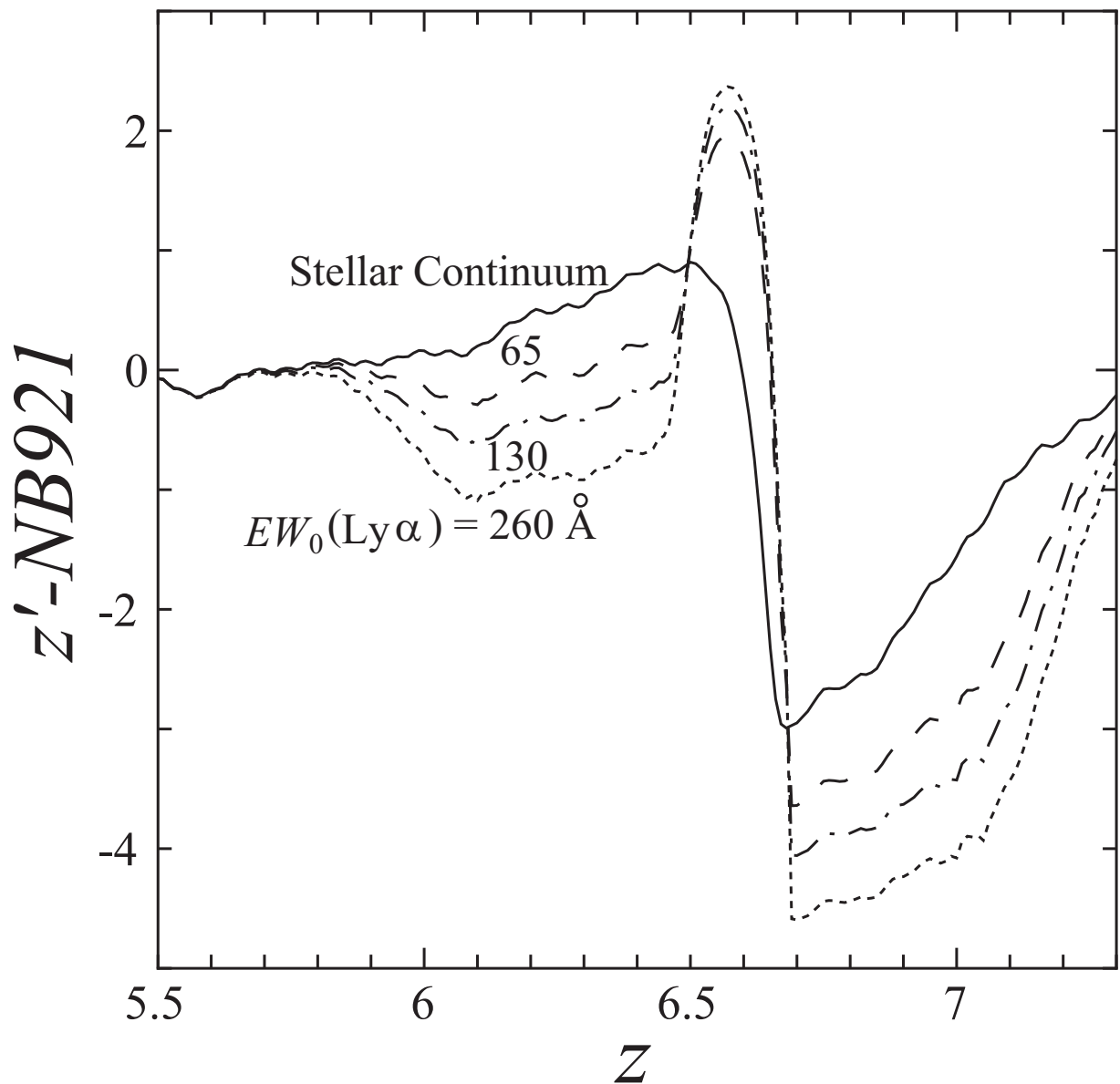}
\caption{
Expected $z^\prime - NB921$ color predicted by the stellar population
synthesis model, as a function of redshift. The solid, dashed,
dash-dotted, and dotted lines denote the stellar SED with Ly$\alpha$ 
whose rest-frame equivalent width is 0, 65, 130, and 260 ${\rm \AA}$,
respectively.
\label{fig5}}
\end{figure}

\clearpage%\newpage

\begin{figure}
\epsscale{1.00}
\plotone{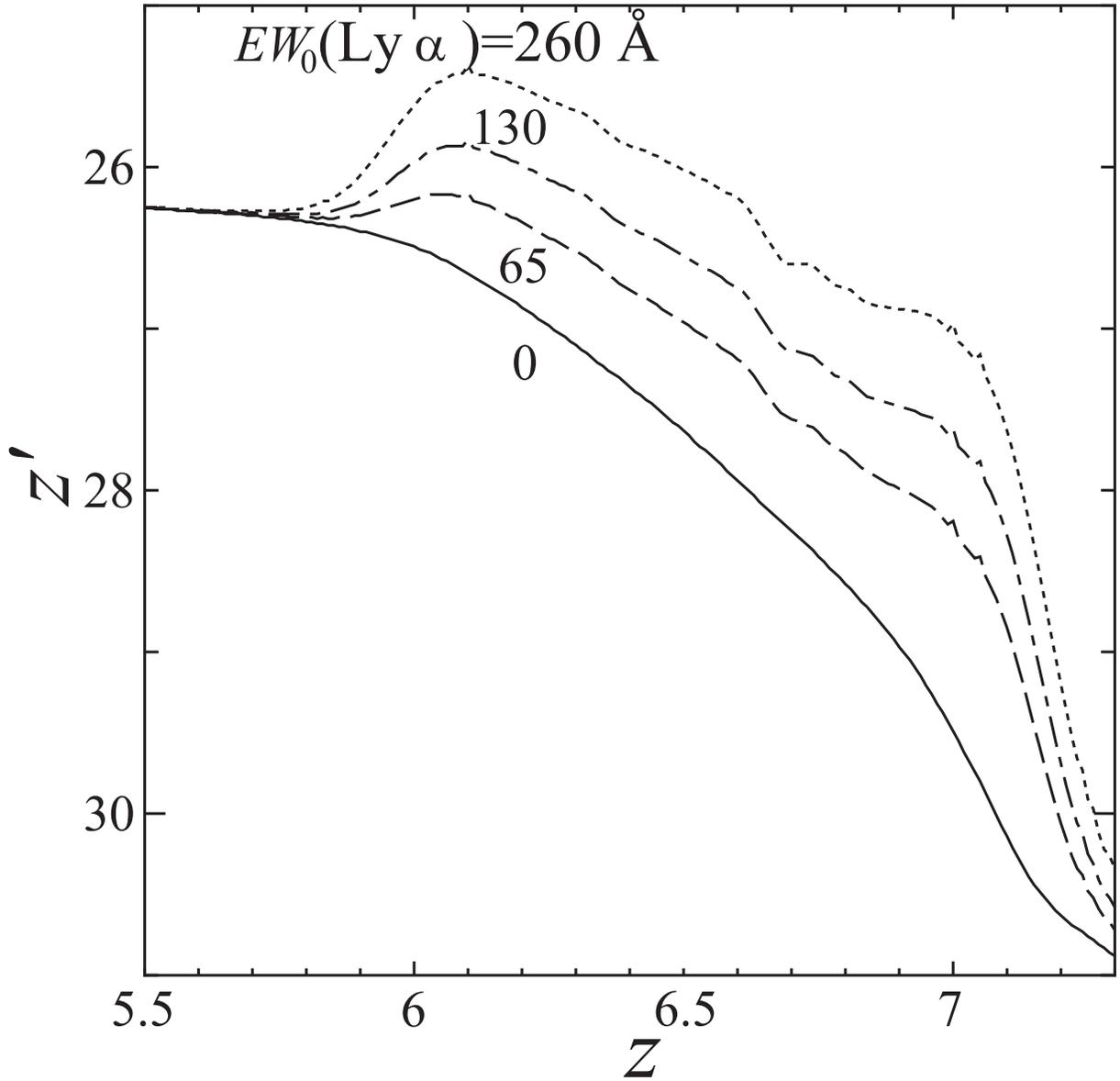}
\caption{
Expected $z^\prime$-band magnitude of galaxies with $M_{1500} = -20.0$
as a function of redshift. The solid, dashed,
dash-dotted, and dotted lines are the same as those in Figure 5.
\label{fig6}}
\end{figure}

\end{document}